\RequirePackage{fix-cm}
\documentclass[smallextended]{svjour3}       
\smartqed  
\usepackage{graphicx}
\usepackage{amsfonts} \usepackage{amsmath}
\usepackage[latin5]{inputenc}

\begin{document}
\newcommand{\aveg}[1]{\left\langle{#1}\right\rangle}
\newcommand{\ket}[1]{\left\vert{#1}\right\rangle}
\newcommand{\bra}[1]{\left\langle{#1}\right\vert}
\newcommand{\nn}{\\\nonumber}

\title{Dynamics of quantum Fisher information in the two-qubit systems constructed from the Yang-Baxter matrices}


\author{Durgun Duran}


\institute{D. Duran \at
Department of Physics, Yozgat Bozok University, Faculty of Science
and Arts, 66100, Yozgat, Turkey. \\
              Tel.: +903542421021\\
              Fax: +903542421022\\
              \email{durgun.duran@bozok.edu.tr}}

\date{Received: date / Accepted: date}

\maketitle

\begin{abstract} By using the quantum Yang-Baxterization approach to the three different Hamiltonians, we investigate the behavior of the quantum Fisher information (QFI) under the
actions of these Hamiltonians on the different two-qubit input states and by estimating the meaningful parameter $\varphi$.  We address the overall estimation properties by evaluating
the QFI for the whole system undergone different unitary evolution. The results show that the behavior of the QFI depends on the choice of the initial states. Choosing the optimal input states
can improve the precision of quantum parameter estimation. On the other hand, we also focus on the dynamical evolution of QFI to distinguish Markovianity and non-Markovianity of the process
by adopting the flow of QFI as the quantitative measure for the information flow. We show that the Hamiltonians constructed with Yang-Baxter matrices influence the dynamics of the system in
the sense of the Markovianity and non-Markovianity. In certain ranges of parameters, we observe that dynamical evolutions of the systems show non-Markovian behavior in which the information
flows from the environment to the system.

\end{abstract}

\section{Introduction}
Parameter estimation plays a crucial role in quantum information theory \cite{Helstrom,Holevo2001,Petz2008}. In this field, determining the value of an unknown parameter that encoded the quantum
system is the main task and enhancing the resolution accuracy is the main goal.  The quantum Cramér-Rao bound (QCRB) provides a fundamental lower bound on the variance of the parameter estimation
and it is proportional to the inverse of the quantum Fisher information (QFI) which is an important figure of merit in quantum parameter estimation theory \cite{Fisher1925,BN2000,Paris2009,Toth2014}.
This bound relates the information obtained about a parameter from measurement outcomes to the parameter estimate uncertainty. The calculation of the QFI for any physical system becomes one of the
central tasks in quantum metrology, although generally, this is difficult. When considering classical probes independently sensing a physical parameter, such as phase or frequency, the maximum attainable
precision follows the standard limit (SL), $1/\sqrt{N}$, where $N$ is the number of probes. In turn, it was shown that quantum entanglement allows one to achieve the so-called Heisenberg limit in precision,
$1/N$, a quadratic improvement as compared to classical approaches \cite{Gio1,Gio2,Gio3}. These precision limits apply to both single-shot protocols as well as protocols utilizing many repetitions.
Still, it remains unclear to what extent such an improvement can be harnessed in practice under non-idealized conditions.

Recently, the Yang-Baxter equations (YBEs) have been introduced to the field of quantum information and quantum computation. In a series of papers, it has been shown that YBEs have a deep connection
with topological quantum computation and entanglement swapping \cite{Kitaev,Kauffman,Franko,Zhang1,Zhang2,Zhang3,Chen1,Chen2,Chen3}. The unitary solution of the braided Yang-Baxter (i.e., the braid
group relation) and unitary solutions of the quantum YBE can often be identified with universal quantum gates \cite{Brylinski,Wang}. This provides a novel way to study quantum entanglement via YBEs.
Usually, a Hamiltonian can be constructed from the unitary $R(\theta, \varphi)$ matrix by the Yang-Baxterization approach. Yang-Baxterization \cite{Jones,Ge1991} has been exploited to derive a Hamiltonian
for the unitary evolution of entangled states. It can be pointed that YBE can be tested in terms of quantum optics \cite{Hu2008}. It is found that any pure two-qudit entangled state can be achieved by
a universal Yang-Baxter matrix (YBM) assisted by local unitary transformations \cite{Chen3}. In recent work, the sudden death of entanglement has been investigated in constructed Yang-Baxter systems (YBEs)
\cite{Hu2010}.

QFI has been applied widely in realizing the different quantum information tasks such as entanglement detection \cite{Li-Luo} and non-Markovian description and determination\cite{Song,Ban}, and flow of QFI
\cite{Lu2010,Zhong2013} that has been used to distinguish Markovian and non-Markovian processes. Exploring open quantum systems from various perspectives has been an intensive research topic in recent years
due to key questions, as well as their important role in the realization of quantum information protocols in real-world applications \cite{Breuer,Alicki,Rivas}. An interesting approach to address open quantum
systems is to investigate the information flow between the components of composite quantum systems, or in particular to explore the exchange of information between the system and its surrounding environment.
In terms of quantum memory effects, dynamical quantum maps are generally divided into two groups: Markovian and non-Markovian maps. Memoryless processes are often recognized as Markovian, where the information
is expected to monotonically flow from the system to the environment. On the other hand, it is rather natural to assume that the backflow of information from the environment to the system is connected to the
presence of memory effects, because in these cases the future states of the system may depend on its past states as a result of the inverse exchange of information.

In this present paper, by using the Yang-Baxterization approach we construct three different Hamiltonians and shall interest in the problem of estimating the parameter that unitary encoded by these 
Hamiltonians. Then, considering two Werner-like states and Bell-diagonal state as input or probe states we firstly investigate the behavior of QFI for the outputs corresponding 
to these inputs and try to achieve optimal conditions saturating the QCRB. On the other hand, we analyze the dynamical behavior of QFI to distinguish Markovianity and non-Markovianity of the process by 
adopting the flow of QFI as the quantitative measure for the information flow evaluating by reduced dynamics of the output state and its physical significance is given.

This study is structured as follows. In Sec. 2 the main traits of QFI and dynamical behavior that will be used in due course are summarized. The YBMs and their properties are carried out in Sec 3.
Hamiltonian models that will be investigated are considered in Sec. 4 where the action of YBM on the Hamiltonians is given. The main results of this work are emphasized in Sec 5 and 6. We end up with
some concluding remarks.

\section{Quantum Fisher Information and Flow of Information}
The QFI indicates the sensitivity of the state to the change of the parameter. Let $\phi$ denote a single parameter to be estimated, for the output state $\rho_{\phi}$
the QFI is generically defined as \cite{Holevo1982,Braunstein}
\begin{eqnarray}
F_{\phi}(\rho_{\phi})=Tr(\rho_{\phi}L_{\phi}^2)=Tr(\partial_{\phi} \rho_{\phi}L_{\phi}),
\end{eqnarray}
where $L_{\phi}$ is symmetric logarithmic derivative(SLD) for the parameter $\phi$, which is a Hermitian operator determined by \cite{Helstrom,Braunstein}
\begin{eqnarray}
\partial_{\phi} \rho_{\phi}=\frac{1}{2}\{\rho_{\phi},L_{\phi}\},
\end{eqnarray}
where $\{\cdot,\cdot \}$ denotes the anticommutator and $\partial_{\phi}\equiv \partial/\partial \phi$.

An essential feature of the QFI is that we can obtain the achievable lower bound of the mean-square error of unbiased estimators for the parameter $\phi$ through the
quantum Cram\'{e}r-Rao (QCR) bound \cite{Cramer1946,Rao}
\begin{eqnarray}
\Delta^2\phi\geq\frac{1}{NF_{\phi}(\rho_\phi)},
\end{eqnarray}
where $N$ is the number of repeated independent measurements. The above inequality defines the principally smallest possible uncertainty in estimation of the parameter. Given the spectral
decomposition of the density operator which is dependent on the parameter $\phi$, $\rho_\phi=\sum_i^s\lambda_i |\psi_i\rangle\langle \psi_i|$ where $\lambda_i$ and $|\psi_i\rangle$ are
respectively the parameter-dependent eigenvalues and eigenstates of $\rho_\phi$ and $s$ is the dimension of the support set of $\rho_\phi$, i.e. $s=dim[supp(\rho_\phi)]$,
then QFI for density matrices with arbitrary ranks can be expressed by \cite{Liu1,Zhang2014,Liu2,Jing,Liu3}
\begin{eqnarray}
F_{\phi}(\rho_\phi) =\sum_{i=1}^{s}\frac{(\partial_\phi\lambda_i)^2}{\lambda_i}
+\sum_{i=1}^{s}4\lambda_i \langle \partial_\phi\psi_i|\partial_\phi\psi_i\rangle-\sum\limits_{\substack{i,j=1 \\ i\neq j}}^s
\frac{8\lambda_i\lambda_j}{\lambda_i+\lambda_j}|\langle \psi_i|\partial_\phi\psi_j\rangle|^2,
\end{eqnarray}
with $\lambda_i+\lambda_j\neq0$. The first term in the right-hand side of Eq. (5) is the classical contribution of QFI whereas the second and third terms can be regarded as the pure quantum
contribution because factor $|\langle \psi_i|\partial_\phi\psi_j\rangle|$ illustrates the quantum coherence between the eigenvectors of $\rho_\phi$.

In the most fundamental parameter estimation task in which the parameter is generated by some unitary dynamics $U=\exp(-i\phi H)$ for some Hamiltonians, $\Delta\phi$ characterizes the estimating
accuracy by any possible measurement made on the quantum state $U\rho U^\dag$ where $\rho$ is the initial probe state. In this situation, the first term in the right-hand side of Eq. (4) vanishes
since the spectrum of the density matrix is unchanged under unitary transformation, no matter the transformation is parameter-dependent or not. Moreover, it is zero for pure states.
In the meantime, with some transformation, Eq. (4) can be rewritten as \cite{Liu3,Boixo,Liu4,Taddei}
\begin{eqnarray}
F_{\phi}(\rho_\phi) =\sum_{i=1}^{s}4\lambda_i \langle \Delta^2\mathcal{H}\rangle_{\psi_i}-\sum\limits_{\substack{i,j=1 \\ i\neq j}}^s \frac{8\lambda_i\lambda_j}{\lambda_i+\lambda_j}|\langle \psi_i|\mathcal{H}|\psi_j\rangle|^2,
\end{eqnarray}
where $\mathcal{H}:=i(\partial_\phi U^{\dagger})U$ is a Hermitian operator since $(\partial_\phi U^{\dagger})U=-U^{\dagger}(\partial_\phi U)$. Here,
\begin{eqnarray}
\langle \Delta^2\mathcal{H}\rangle_{\psi_i}=\langle \psi_i|\mathcal{H}^2 |\psi_i\rangle-|\langle \psi_i|\mathcal{H} |\psi_i\rangle|^2
\end{eqnarray}
is the variance of $\mathcal{H}$ on the ith eigenstate of the input state $\rho$.

For the dynamical behavior of QFI, we introduce the QFI flow, which is defined as the change rate of the QFI by \cite{Lu2010}
\begin{eqnarray}
\mathcal{I_\phi}(\rho_\phi)=\frac{\partial F_{\phi}(\rho_\phi)}{\partial\mathrm{t}}.
\end{eqnarray}
It is well-known that $\mathcal{I_\phi}(\rho_\phi)<0$ for some $t$ represents the information flow from system to the environment which defines the Markovian regime and the QFI is monotonically
decreasing under Markovian dynamics, as it cannot increase under completely positive maps while $\mathcal{I_\phi}(\rho_\phi)>0$  corresponds to the non-Markovian regime where information flow
is from the environment to the system \cite{Lu2010,Zhong2013,Lin2008,Abdel-Khalek,Haikka,Hao1,Hao2}. In this situation, it can be called that QFI witnesses the non-Markovianity of the dynamics of the system.

Especially, this witness of non-Markovianity may be relevant in the context of quantum parameter estimation. Specifically, the error (variance) of any (unbiased) estimation of the
parameter $\phi$ is related to the QFI through the QCRB given by Eq. (3). Thus, an increment in $F_{\phi}$ could be linked with an increment of information about the parameter $\phi$.
Nevertheless, note that the QFI provides just a lower bound to the error on $\phi$, and in fact there are cases where this bound is not achievable.

The QFI and the dynamical behavior of QFI are the primary focus of this work and we further discuss its role in quantifying the precision of estimation in more detail.

\section{Yang-Baxter Matrices}
A class of invariants of knots and links called quantum invariants can be constructed by using representations of the Artin braid group, and more specifically by using solutions to the
YBE \cite{Yang,Baxter}, first discovered concerning $1+1$ dimensional quantum field theory, and two-dimensional models in statistical mechanics. Braiding operators feature in constructing
representations of the Artin braid group, and in the construction of invariants of knots and links. A key concept in the construction of quantum link invariants is the association of a
Yang-Baxter operator $R$ to each elementary crossing in a link diagram. The operator $R$ is a linear mapping \cite{Kauffman} $R:V\otimes V \rightarrow V\otimes V$ defined on the two-fold
tensor product of a vector space $V$, generalizing the permutation of the factors (i.e., generalizing a swap gate when $V$ represents one qubit). Such transformations are not necessarily
unitary in topological applications. It is useful to understand when they can be replaced by unitary transformations for quantum computing. Such unitary $R$-matrices can be used to make
unitary representations of the Artin braid group.

A solution to the YBE, as described above is a matrix $R$, regarded as a mapping of a two-fold tensor product of a vector space $V\otimes V$ to itself that satisfies the equation
\begin{eqnarray}
(R\otimes \mathbb{I})(\mathbb{I}\otimes R)(R\otimes \mathbb{I})=(\mathbb{I}\otimes R)(R\otimes \mathbb{I})(\mathbb{I}\otimes R),
\end{eqnarray}
where $\mathbb{I}$ is the identity operator.

In this paper, we need to study solutions of the YBE that are unitary to relate quantum computing and quantum entanglement. Then the $R$ matrix can be seen either as a braiding matrix or as
a quantum gate in a quantum computer.

The unitary $R$-matrix satisfies the YBE
\begin{eqnarray}
R_i(\mu)R_{i+1}(\mu+\nu)R_i(\nu)=R_{i+1}(\nu)R_i(\mu+\nu)R_{i+1}(\mu),
\end{eqnarray}
or
\begin{eqnarray}
R_{i}(\mu)R_{i+1}\left(\frac{\mu+\nu}{1+\beta^2\mu\nu}\right)R_{i}(\nu)=R_{i+1}(\nu)R_{i}\left(\frac{\mu+\nu}{1+\beta^2\mu\nu}\right)R_{i+1}(\mu),
\end{eqnarray}
where $\beta =-i/c$ (c is the velocity of light)\cite{Jimbo1989}, $\mu$ and $\nu$  are the parameters which usually range over the real numbers $\mathbb{R}$ in the case of an additive parameter,
or over positive real numbers $\mathbb{R^+}$ in the case of a multiplicative parameter. It is worth noting that the four-dimensional YBE Eqs. (9) and (10) admit the Temperly-Lieb algebra (TLA)
\cite{Temperley,Hu2007}. Actually the rational solution of the YBE, $R(\mu)$ can be written in terms of a unitary transformation $U$ in the following way: $R(\mu) = a(\mu)\mathbb{I} + b(\mu)U$,
with $U$ satisfying the TLA
\begin{eqnarray}
U_iU_{i+1}U_i=U_i,\quad U_{i}^2=dU_i,\quad U_iU_{j}=U_{j}U_i
\end{eqnarray}
for $|i-j|\geq2$, where $d$ is the single loop in the knot theory which does not depend on the sites of the lattices. When $d=2$, the Hermitian matrix $U$ has forms as follows
\begin{eqnarray}
U_{1}=\left( {\begin{array}{cccc}
1 & 0 & 0 & e^{i\varphi}\\
0 & 0 & 0 & 0\\
0 & 0 & 0 & 0\\
e^{-i\varphi} & 0 & 0 & 1
\end{array}}
\right),\quad
U_{2}=\left( {\begin{array}{cccc}
0 & 0 & 0 & 0\\
0 & 1 & e^{i\varphi} & 0\\
0 & e^{-i\varphi} & 1 & 0\\
0 & 0 & 0 & 0
\end{array}}
\right),
\end{eqnarray}
When $d=\sqrt{2}$, the Hermitian matrix U takes the form
\begin{eqnarray}
U_{3}=\frac{1}{\sqrt{2}}\left( {\begin{array}{cccc}
1 & 0 & 0 & e^{i\varphi}\\
0 & 1 & i\varepsilon & 0\\
0 & -i\varepsilon & 1 & 0\\
e^{-i\varphi} & 0 & 0 & 1
\end{array}}
\right)
\end{eqnarray}
where $\varphi$ is real and $\varepsilon=\pm1$.

Three unitary matrices $R_{i,i+1}(\theta, \varphi)$ are obtained by the Yang-Baxterization approach \cite{Jones,Ge1991} according to the above $U$ matrices as follows,
\begin{subequations}
\begin{align}
R_{i,i+1}^{(1)}(\theta, \varphi)=&\left(\cos\frac{\theta}{2}+\frac{i}{2}\sin\frac{\theta}{2}\right)\mathbb{I}_i \mathbb{I}_{i+1}-2i\sin\frac{\theta}{2} S_i^z S_{i+1}^z \nonumber\\
&-i\sin\frac{\theta}{2}\left(e^{i\varphi}S_i^+ S_{i+1}^{+}+e^{-i\varphi}S_i^- S_{i+1}^-\right),\\
R_{i,i+1}^{(2)}(\theta, \varphi)=&\left(\cos\frac{\theta}{2}+\frac{i}{2}\sin\frac{\theta}{2}\right)\mathbb{I}_i \mathbb{I}_{i+1}+2i\sin\frac{\theta}{2} S_i^z S_{i+1}^z\nonumber\\
&-i\sin\frac{\theta}{2}\left(e^{i\varphi}S_i^+ S_{i+1}^{-}+e^{-i\varphi}S_i^- S_{i+1}^+\right),\\
R_{i,i+1}^{(3)}(\theta, \varphi)=&-\cos\frac{\theta}{2}\mathbb{I}_i \mathbb{I}_{i+1}-i\sin\frac{\theta}{2}(e^{i\varphi}S_i^+ S_{i+1}^{+}+e^{-i\varphi} S_i^{-} S_{i+1}^{-})\nonumber\\
&+\varepsilon\sin\frac{\theta}{2}(S_i^+ S_{i+1}^{-}-S_i^- S_{i+1}^+),
\end{align}
\end{subequations}
where $S_i^z$ is the spin operators for the ith particle and $S_i^{\pm} = S_i^{x} \pm iS_i^{y}$ are raising and lowering operators respectively for the ith particle. The parameter $\theta$
appearing in Eqs. (14a) and (14b) is related to $\mu$ as $\cos\theta=(1-\mu^2)/(1+\mu^2)$. In Eq. (14c), the relation of $\theta$ and $\mu$ can be written as $\cos\theta=1/\cosh\mu$. Note that
solutions of the YBE for $d=2$ are given by meromorphic functions of $\mu$ whereas for $d\neq2$ by trigonometric functions. The difference of two $\theta$ and $\mu$ relations come from this
property of YBE.

\section{Dynamical Models}
Consider a system of two spin-$1/2$ particles (particle $1$ and $2$) or nearest spin-spin interaction described by an initial Hamiltonian $H_0$ \cite{Hu2010,Sun2009}
\begin{eqnarray}
H_0=\mu_1 S_1^z+\mu_2 S_2^z+gS_1^zS_2^z,
\end{eqnarray}
where $\mu_i$ represents external magnetic field and $g$ is the coupling constant of z-component of two neighboring spins. For convenience of calculations,
we introduce two parameters $B=(\mu_1+\mu_2)/2$ and $J=(\mu_1-\mu_2)/2$. Taking into account the Schr\"{o}dinger equation
\begin{eqnarray}
i\hbar\frac{\partial}{\partial t}|\Psi(\theta,\varphi)\rangle=H(\theta,\varphi)|\Psi(\theta,\varphi)\rangle
\end{eqnarray}
and $|\Psi(\theta,\varphi)\rangle=R(\theta, \varphi)|\Psi_0\rangle$ one obtains
\begin{eqnarray}
i\hbar\frac{\partial}{\partial t}R(\theta, \varphi)|\Psi_0\rangle=H(\theta,\varphi)R(\theta, \varphi)|\Psi_0\rangle
\end{eqnarray}
where $|\Psi_0\rangle$ is the eigenstate of $H_0$. Let real parameters $\theta$ and $\varphi$ be time-independent, one can get a Hamiltonian through the unitary transformation
$R(\theta, \varphi)$ as $H(\theta,\varphi)=R(\theta, \varphi)H_0 R^{-1}(\theta, \varphi)$. Now, three Hamiltonians are obtained from Eqs. (14) as follows \cite{Hu2010}
\begin{subequations}
\begin{align}
    H_1(\theta, \varphi)=&B\cos\theta(S_1^z+S_2^z)+J(S_1^z-S_2^z)+gS_1^zS_2^z\nonumber\\
    &+iB\sin\theta\left(e^{i\varphi}S_1^+ S_2^{+}-e^{-i\varphi}S_1^- S_2^-\right),\\
    H_2(\theta, \varphi)=&B(S_1^z+S_2^z)+J\cos\theta(S_1^z-S_2^z)+gS_1^zS_2^z\nonumber\\
    &+iJ\sin\theta\left(e^{i\varphi}S_1^+ S_2^{-}-e^{-i\varphi}S_1^- S_2^+\right),\\
    H_3(\theta, \varphi)=&B\cos\theta(S_1^z+S_2^z)-iB\sin\theta\left(e^{i\varphi}S_1^+ S_2^{+}-e^{-i\varphi}S_1^{-} S_2^{-}\right)\nonumber\\
    &+gS_1^zS_2^z+J\cos\theta(S_1^z-S_2^z)+\varepsilon J\sin\theta\left(S_1^+ S_2^{-}+S_1^- S_2^+\right).
\end{align}
\end{subequations}

Specifically, for $\varphi=-\pi/2$ we find that the second model is the 2-qubit anisotropic Heisenberg XXZ model under an inhomogeneous magnetic field, and
the third model is the 2-qubit anisotropic Heisenberg XYZ model in an inhomogeneous magnetic field.

\section{Quantum parameter estimation in Yang-Baxter Systems}
In this section, we investigate the dynamics of QFI for three Hamiltonians under the adjoint action of unitary YBE $R(\theta, \varphi)$ on the bipartite two-qubit input states.
We first fix two Werner-like states and later consider the Bell-diagonal states as an input state.

For convenience, we set $\theta=\pi/2$, henceforward. So, we can rewrite Hamiltonians in Eqs. (18) as follows ($\varepsilon=1$)

\begin{subequations}
\begin{align}
H_1( \varphi)=&J(S_1^z-S_2^z)+gS_1^zS_2^z+iB\left(e^{i\varphi}S_1^+ S_2^{+}-e^{-i\varphi}S_1^- S_2^-\right),\\
H_2(\varphi)=&B(S_1^z+S_2^z)+gS_1^zS_2^z+iJ\left(e^{i\varphi}S_1^+ S_2^{-}-e^{-i\varphi}S_1^- S_2^+\right),\\
H_3(\varphi)=&J\left(S_1^+ S_2^{-}+S_1^- S_2^+\right)+gS_1^zS_2^z-iB\left(e^{i\varphi}S_1^+ S_2^{+}-e^{-i\varphi}S_1^{-} S_2^{-}\right).
\end{align}
\end{subequations}
For the action of Hamiltonian $H_3$ since the spectrum of output is same as that of the first one it is easy to find that in the third YBS we get the same result
as the first one for all probe states and hereafter, it is not reported for the next sections.

\subsection{Action of $H_1$ to the Initial Werner-like States and Behavior of QFI}

For a general initial input states $\rho$, the output $\sigma$ under the unitary time evolution or unitary adjoint action $ad_{U}(\cdot)=U(\cdot)U^{\dagger}$ is found to be
\begin{eqnarray}
\sigma=U\rho U^{\dagger}=e^{-itH}\rho e^{itH}.
\end{eqnarray}

Firstly, we investigate the behavior of QFI for the outputs under action of Hamiltonian $H_1$ on two initial Werner-like states that denoted $\rho_{AB}^{(j)}$ with $j=1,2$.
We first fix the two-qubit probe state (Werner state) to be
\begin{eqnarray}
\rho_{AB}^{(1)}= (1-p)\frac{\mathbb{I}}{4}+p|\beta_{00}\rangle \langle\beta_{00}|,
\end{eqnarray}
where $p\in[0,1]$, $\mathbb{I}$ is the $4\times4$ identity matrix and the mnemonic notation $|\beta_{xy}\rangle$ can be understood via the equations
\begin{eqnarray}
|\beta_{xy}\rangle\equiv \frac{1}{\sqrt{2}}(|0,y\rangle+(-1)^x|1,\bar{y}\rangle)
\end{eqnarray}
in the standard two-qubit computational basis $\{|00\rangle,|01\rangle, |10\rangle, |11\rangle\}$. Here $\bar{y}$ is the negation of $y$ \cite{NC}.

From here on under the action of Hamiltonians $H_i (i=1,2)$
the QFIs of the outputs $\sigma_{AB}^{(j)} (j=1,2)$ will be respectively denoted by $F^{(i)}_{\varphi}(\sigma_{AB}^{(j)})$.

In general, the spectrum of output is unchanged since the evolution is unitary. Now, we can calculate the QFI concerning to the estimated parameter $\varphi$ with the help of Eq.(5).
The QFI for the output state $\sigma_{AB}^{(1)}$ under the action of Hamiltonian $H_1$ on the input state $\rho_{AB}^{(1)}$ is obtained as
\begin{eqnarray}
F^{(1)}_{\varphi}(\sigma_{AB}^{(1)})=\frac{8p^4}{1+p}\sin^2(Bt) \left[1-\cos^2\varphi\cos^2(Bt) \right].
\end{eqnarray}

As a second example, we consider a different two-qubit Werner-like state
\begin{eqnarray}
\rho_{AB}^{(2)}=p|\beta_{11}\rangle\langle \beta_{11}|
+\frac{1-p}{2}(|\beta_{01}\rangle\langle \beta_{01}|+|\beta_{00}\rangle\langle \beta_{00}|)
\end{eqnarray}
as an input state in the standard basis. In this situation, we can again calculate the QFI for output state $\sigma_{AB}^{(2)}$ under the action of Hamiltonian $H_1$ on the input $\rho_{AB}^{(2)}$ as follows
\begin{eqnarray}
F^{(1)}_{\varphi}(\sigma_{AB}^{(2)})= 2(1-p)\sin^2(Bt) \left[1-\cos^2\varphi\cos^2(Bt)\right].
\end{eqnarray}

\begin{figure}[!btp]
\centering
\includegraphics[width=12cm]{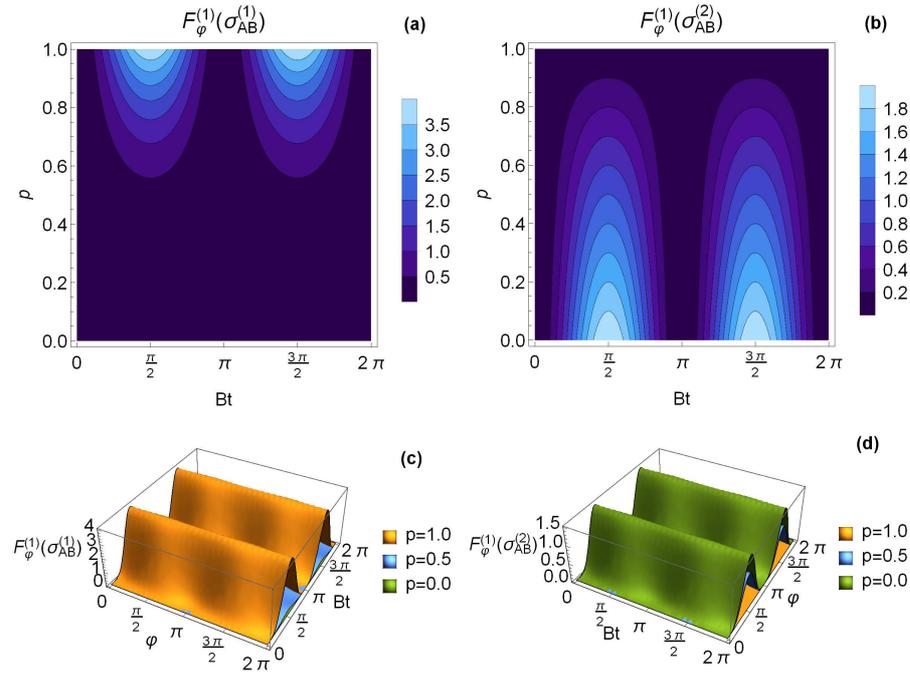}
\caption{(color online) Plots of QFIs for the outputs $\sigma_{AB}^{(1)}$ and $\sigma_{AB}^{(2)}$ under the action of $H_1$ on the these two inputs.
In (a) and (b), we take $\varphi=\pi/2$ and in this situation QFIs have quite different behaviors. Similarly, in (c) and (d) maximizing the QFI depends entirely on different parameters.
It depends on only choice of the parameter $Bt$ in (c) while in (d) this depends on the choice of the estimated parameter $\varphi$.}
\end{figure}

Figure 1 displays QFIs of the outputs $\sigma_{AB}^{(1)}$ and $\sigma_{AB}^{(2)}$ under the action of $H_1$ on the these two inputs. In Fig. 1(a) and (b), we depict the QFI as a function of $Bt$ and $p$ for
the fixed value of estimated parameter $\varphi=\pi/2$. It is noted that $F^{(1)}_{\varphi}(\sigma_{AB}^{(1)})$ can be written in terms of $F^{(1)}_{\varphi}(\sigma_{AB}^{(2)})$ as
$F^{(1)}_{\varphi}(\sigma_{AB}^{(1)})=4p^4 F^{(1)}_{\varphi}(\sigma_{AB}^{(2)})/(1-p^2)$. From Fig. 1(a), QFI increases with the increasing values of parameter $p$, especially in the region of $p>1/2$,
and naturally attains the maximum value for $Bt=k\pi/2$ with any odd $k$ and $p=1$ where the input state is maximally entangled, namely Bell state. Besides all these, it decreases with decreasing values
of the estimated parameter $\varphi$ and also attains its maximum value for the estimated parameter $\varphi=\pi/2$ in which the behavior of QFI is depicted for this value in Fig. (1).
Also, it shows a periodic behavior according to $Bt$. On the other hand, QFI has the opposite behavior in Fig. 1(b) compared to Fig. 1(a). It takes place its maximum value in the small values of $p$,
especially $p=0$ in which the probe state is reduced to a mixture of two Bell states with equal probability and $Bt=k\pi/2$.

In Fig. 1(c) and (d), we give the plots of the QFI versus the $Bt$ and the estimated parameter $\varphi$ for the different values of the initial state parameter $p$. In Fig. 1(c), the maximizing of the QFI strictly
depends on the choice of the parameter $Bt$ for all values of the $\varphi$ whereas it depends on the choice of the estimated parameter $\varphi$. They have a quite opposite behavior. Additionally, in Fig. 1(c) and (d)
QFI vanishes for the values of the parameter $p=0$ and $p=1$, respectively. We conclude that the maximizing of QFI has a significant connection with the choice of the initial state.

\subsection{Action of $H_2$ to the Initial Werner-like States and Behavior of QFI}

In this case, the QFI for the output state $\sigma_{AB}^{(1)}$ under the action of Hamiltonian $H_2$ on the input $\rho_{AB}^{(1)}$ vanishes, $F^{(2)}_{\varphi}(\sigma_{AB}^{(1)})=0$.

For the second probe state $\rho_{AB}^{(2)}$, the QFI can be calculated as follows
\begin{eqnarray}
F^{(2)}_{\varphi}(\sigma_{AB}^{(2)})= \frac{2(1-3p)^2}{1+p} \sin^2(Jt)\left[1-\cos^2\varphi\cos^2(Jt)\right].
\end{eqnarray}

\begin{figure}[!btp]
\centering
\includegraphics[width=12cm]{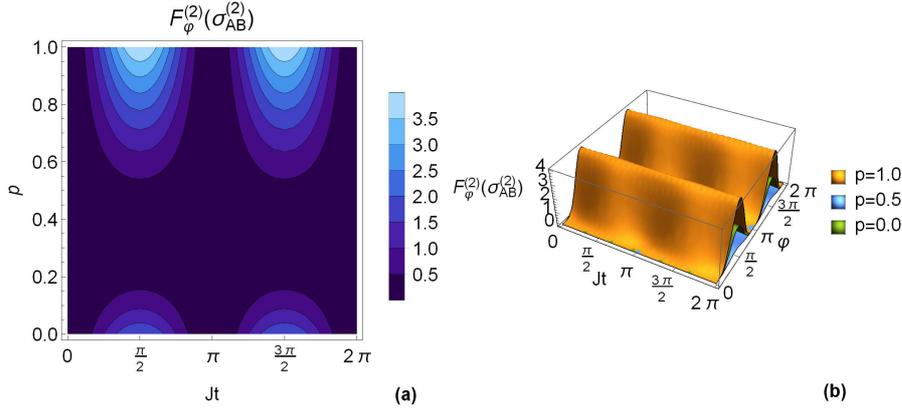}
\caption{(color online) Plots of the QFI given by Eq. (26) under action of the Hamiltonian $H_2$ on the input state given by Eq. (24) versus (a) $p$ and $Jt$ with $\varphi=\pi/2$, (b) $\varphi$ and $Jt$
for the fixed values of $p$.}
\end{figure}
In Fig. 2, we give the plots of QFI as a function of (a) $p$ and $Jt$ for the value of $\varphi=\pi/2$ and (b) $\varphi$ and $Jt$ for the different values of $p$. From Fig. 2(a),  we can say that
QFI generally increases for the increasing values of $p$. However, it can be observed that some enhancements in QFI are obtained for the small values of $p$. Especially, when $Jt=k\pi/2$ with any odd $k$
QFI attains its maximum values for $p=1$ in which input state corresponds to maximally entangled two-qubit state, namely Bell state. For the intermediate values of $p$ it vanishes independently
of the value of $Jt$.

On the other hand, in Fig. 2(b), QFI is maximized for the values of the estimated parameter $\varphi=k\pi/2$ irrespective of the parameter $Jt$. Because of the above observations,
these imply that QFI can be relatively enhanced by adjusting the parameters $\varphi$, $p$ and $Jt$. By adopting the QCRB as a figure of merit, these enhancements can be clearly shown for optimal
parameter estimation. As a result, we can conclude that the QFI not only depends on the choice of the initial condition but also has a connection with the actions of the different Hamiltonians.

\subsection{Actions of Hamiltonians to Initial Bell-diagonal States}
More generally, let us now consider two-qubit Bell-diagonal states as input states \cite{Luo}
\begin{eqnarray}
\rho_{AB}=\frac{1}{4}\Big(\mathbb{I}\otimes\mathbb{I}+\sum_{i=1}^3c_i\sigma_i\otimes \sigma_i\big),
\end{eqnarray}
where matrices $\sigma_i$ are the Pauli spin matrices and real numbers $c_i$ fulfill the following conditions
\begin{subequations}
\begin{align}
0\leq\frac{1}{4}(1-c_1-c_2-c_3)\leq1,\\
0\leq\frac{1}{4}(1-c_1+c_2+c_3)\leq1,\\
0\leq\frac{1}{4}(1+c_1-c_2+c_3)\leq1,\\
0\leq\frac{1}{4}(1+c_1+c_2-c_3)\leq1,
\end{align}
\end{subequations}
where $c_1, c_2, c_3\in[-1,1]$.

Since the evolution of the system under the action of Hamiltonian is unitary, the spectrum of the output state $\sigma_{AB}$ is unchanged and is given by terms between the inequalities
in the Eqs. (28). We denote the QFIs corresponding to the output $\sigma_{AB}$ under the action of the two Hamiltonians $H_i (i=1,2)$ on the input state $\rho_{AB}$ as
$F_{\varphi}^{(1)}(\sigma_{AB})$ and $F_{\varphi}^{(2)}(\sigma_{AB})$, respectively. So, QFIs for the actions of $H_1$ and $H_2$ are calculated as follows
\begin{eqnarray}
F_{\varphi}^{(1)}(\sigma_{AB})&=&\frac{(c_1-c_2)^2}{2(1+c_3)}\sin^2(Bt)[1-\cos^2\varphi \cos^2(Bt)],\\
F_{\varphi}^{(2)}(\sigma_{AB})&=&\frac{(c_1-c_2)^2}{2(1+c_3)}\sin^2(Jt)[1-\cos^2\varphi \cos^2(Jt)],
\end{eqnarray}
respectively. It is noted that we get the similar behaviors for these YBSs. The only thing that has changed is $B\rightarrow J$ transformation.

\begin{figure}[!btp]
\centering
\includegraphics[width=7cm]{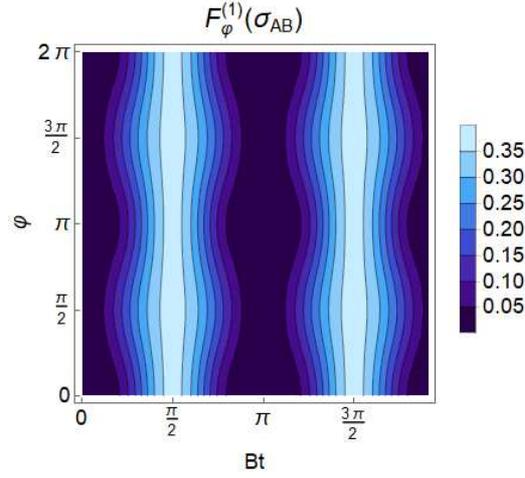}
\caption{(color online) Plot of the QFI given by Eq. (29) under action of the Hamiltonian $H_1$ with respect to $Bt$ and $p$ with $c_1=0.9,c_2=0$ and $c_3=0.1$.}
\end{figure}

In Fig. 3, we give the only plot of QFI of the output $\sigma_{AB}$ under the action of $H_1$ as a function of $Bt$ and $\varphi$ for the values of $c_1=0.9,c_2=0$ and $c_3=0.1$. For both QFIs
given by Eq. (29) and (30), the maximum values are reached at $Bt=m\pi/2$ ($m$ is an odd number) irrespective of $\varphi$. Since the QFIs have the same behavior, the plot of Eq. (30) is not
depicted here. On the other hand, QFI vanishes for the intermediate values of $Bt$. Evidently, it can be enhanced by the appropriate choice of parameters $c_1,c_2$ and $c_3$.

\section{Dynamical Behavior of QFI}
In this section, we consider the information flow for the reduced dynamics of the outputs under the actions of the Hamiltonians $H_1$ and $H_2$ on the input states given by
Eqs. (21), (24) and (27). Here, the second particle $B$ can be considered to act as the environment.

Firstly, we investigate the dynamics of QFIs of the reduced density matrices of the outputs under the action of $H_1$ on the input states $\rho^{(1)}_{AB}$ and $\rho^{(2)}_{AB}$.
From Eqs. (4) and (7), QFI and the flow of QFI can respectively be calculated as follows
\begin{subequations}
\begin{align}
F_{\varphi}^{(1)}(\sigma^{(1)}_{A})=&\frac{p^2\sin^2\varphi}{\csc^2(2Bt)-p^2\cos^2\varphi},\\
\mathcal{I}^{(1)}_\varphi(\sigma^{(1)}_{A})=&\frac{4Bp^2 \sin^2\varphi\cot(2Bt) \csc^2(2Bt)}{[\csc^2(2Bt)-p^2 \cos^2\varphi]^2}.
\end{align}
\end{subequations}

\begin{figure}[!btp]
\centering
\includegraphics[width=11cm]{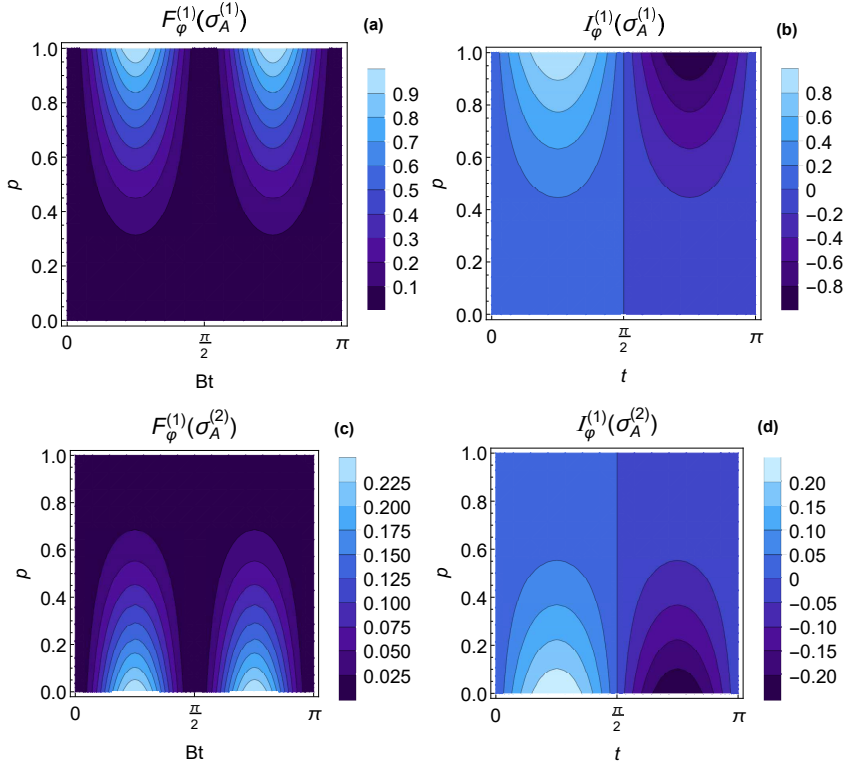}
\caption{(color online) Plots of the QFI and flow of QFI given by Eq. (31) and (32) for the reduced density matrices $\sigma^{(1)}_{A}$ and $\sigma^{(2)}_{A}$ under the action of $H_1$ on the
all input states. For all plots we take $\varphi=\pi/2$ in which the QFI takes place its maximum value. For (c) and (d), we have chosen $2B=1$.}
\end{figure}

Similarly, for the second input state $\rho^{(2)}_{AB}$, QFI and the flow of information for the reduced density matrix $\sigma^{(2)}_{A}$ of the output $\sigma^{(2)}_{AB}$ are obtained
\begin{subequations}
\begin{align}
F_{\varphi}^{(1)}(\sigma^{(2)}_{A})=&\frac{x\sin^2\varphi}{4-x\cos^2\varphi },\\
\mathcal{I}^{(1)}_\varphi(\sigma^{(2)}_{A})=&\frac{16B\sin^2\varphi\sqrt{x[(1-p)^2-x]}}{(4-x\cos^2\varphi)^2},
\end{align}
\end{subequations}
respectively. Here, $x=(1-p)^2\sin^2(2Bt)$.

Plots of the QFIs and the flows of QFIs given by Eq. (31) and (32) are depicted in Fig. (4). From Fig. 4(a) and (c), QFI has the same behavior as that of Fig. (1). It should be noted that
the maximum value of QFI strictly depends on the choice of the initial state where the same Hamiltonian acts on the different initial states. On the other hand, from Fig. 4(b) and (d)
$\mathcal{I}^{(1)}_\varphi$ takes place the negative values for the values of $k\pi/2<t<l\pi$ where $k$ is an odd number and $l$ is an even number. Therefore, the dynamical evolution of the
system is Markovian because $\mathcal{I}^{(1)}_\varphi<0$ indicates that the energy and information flow out from the atom or system and $\mathcal{I}^{(1)}_\varphi>0$ represents the energy and
information flow in the atom from the environment. So, when $\mathcal{I}^{(1)}_\varphi>0$ the dynamics of the system is non-Markovian. Also, for the negative values of $\mathcal{I}^{(1)}_\varphi$
when it decreases, QFI increases.

Secondly, for the action of the Hamiltonian $H_2$ on the two input states, similar behaviors are observed with the previous case. It is noted that for the first input state since the QFI of the output
vanishes the QFI and flow of QFI for the reduced density matrix is zero. So, under the action of $H_2$ on the second input states, QFI and its flow are found to be as
\begin{subequations}
\begin{align}
F_{\varphi}^{(2)}(\sigma^{(2)}_{A})=&\frac{y\sin^2 \varphi}{4-y\cos^2\varphi},\\
\mathcal{I}^{(2)}_\varphi(\sigma^{(2)}_{A})=&\frac{16J\sin^2 \varphi \sqrt{y[(1-3p)^2-y]}}{(4-y\cos^2\varphi)^2},
\end{align}
\end{subequations}
respectively. Here $y=(1-3p)^2\sin^2(2Jt)$.

\begin{figure}[!btp]
\centering
\includegraphics[width=11cm]{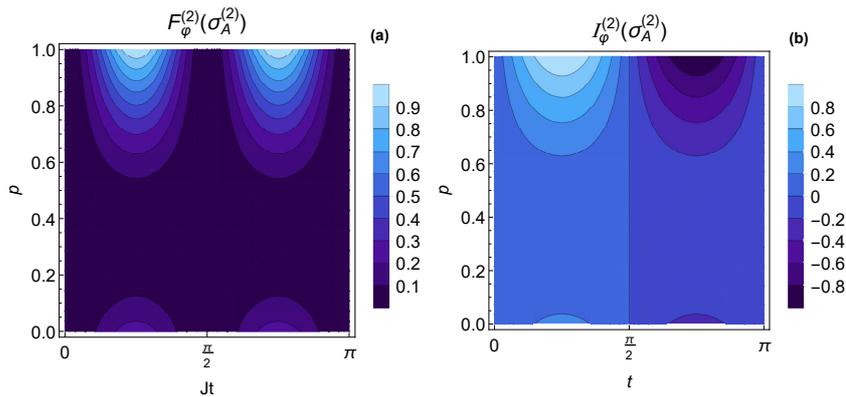}
\caption{(color online) Plots of the QFI and flow of QFI for the action of the Hamiltonian $H_2$ with respect to $Jt$ and $t$, respectively.
For both plots we take $\varphi=\pi/2$ in which the QFI takes place its maximum value and we have chosen $2J=1$ for the second plot.}
\end{figure}

Plots the QFI and flow of information are depicted in Fig. (5) versus $Jt$ and $t$, respectively. Analogous with the previous case, the flow of QFI has the same behavior for the values of parameters.
However, it is noted that in Fig. 4(c) QFI attains its maximum value for $p=1$ and $Bt=n\pi/4 (n=1,3,...)$ whereas it takes place the maximum at $p=1$ and the values of parameter $Jt=n\pi/4$
in Fig. 5(a). It can also be seen from Fig. 5(b) information flows from the environment to the system in which the dynamical behavior of the system represents the non-Markovian regime.

Finally, we investigate the dynamics of QFI for Bell-diagonal input states under the action of the Hamiltonians $H_1$ and $H_2$. The QFIs and again the flows of QFIs for the two YBSs are explicitly
calculated as follows
\begin{subequations}
\begin{align}
F_{\varphi}^{(1)}(\sigma_{A})=&\frac{(c_1-c_2)^2}{4\csc^2\varphi\csc^2(2Bt)-(c_1-c_2)^2\cot^2\varphi},\\
\mathcal{I}^{(1)}_\varphi(\sigma_{A})=&\frac{8B(c_1-c_2)^2 \csc^2\varphi \sin(4Bt) \csc^4(2Bt)}{[4\csc^2\varphi\csc^2(2Bt)-(c_1-c_2)^2\cot^2\varphi]^2},
\end{align}
\end{subequations}
and
\begin{subequations}
\begin{align}
F_{\varphi}^{(2)}(\sigma_{A})=&\frac{(c_1+c_2)^2}{4\csc^2\varphi\csc^2(2Jt)-(c_1+c_2)^2\cot^2\varphi},\\
\mathcal{I}^{(2)}_\varphi(\sigma_{A})=&\frac{16J(c_1+c_2)^2 \csc^2\varphi \cot(2Jt) \csc^2(2Jt)}{[4\csc^2\varphi\csc^2(2Jt)-(c_1+c_2)^2\cot^2\varphi]^2},
\end{align}
\end{subequations}
respectively.
\begin{figure}[!btp]
\centering
\includegraphics[width=11cm]{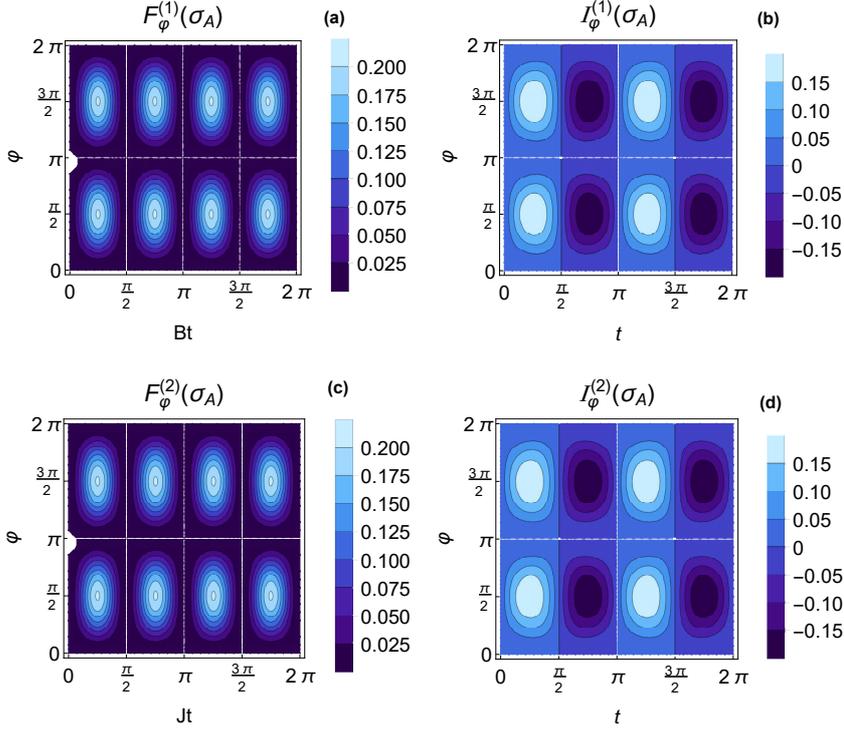}
\caption{(color online) Plots of the QFI and flow of QFI for initial Bell-diagonal state under the action of first two Yang-Baxter systems.
For all plots we have chosen the parameters $c_1=0.9, c_2=0, c_3=0.1$, $2B=1$ and $2J=1$ for (b) and (d), respectively.}
\end{figure}

Plots the QFI and information flow for the reduced dynamics under the action of Hamiltonians $H_1$ and $H_2$ on the Bell-diagonal input state given by Eq. (27) are shown in Fig. 6 as functions of the
parameters $\varphi$, $Bt$, $Jt$ and $t$. In Fig. 6(a) and (c), QFIs have a similar behavior since both of them have the same form depending on the selected values of the parameters. For both figures,
QFI attains its maximum values at $Bt (Jt)=k\pi/4$ and $\varphi=k\pi/2$ where k is an odd number. It can be also seen that it takes place the minimum values for $Bt (Jt)=n\pi/2$ and $\varphi=n\pi$ with
all integer values of $n$.

On the other hand, for some values of $t$ flow of information $\mathcal{I}^{(j)}_\varphi(\sigma_{A}) (j=1,2)$ takes the positive values that describe the non-Markovian evolution similar to the previous
cases in Fig. 6(b) and (d). Evidently, it can be said that under the action of both YBSs, due to the memory and feedback effect of the non-Markovian environment, energy and information flow from the
environment to the system, accounting for the QFI of revival.

\section{Concluding Remarks}
In this paper, some Hamiltonians have been constructed by the unitary YBMs $R_{i,i+1}(\theta, \varphi)$ from a Hamiltonian $H_0$ describing the nearest spin-spin interaction where parameters
$\theta$ and $\varphi$ are time-independent. Firstly, we have studied the behavior of QFI quantifying the information content of a quantum state concerning a given observable for two-qubit systems
under the actions of unitary Yang-Baxter channels or Yang-Baxterization approach. Our results clearly indicate that QFI shows different behavior for different input states and different YBSs. 
Under the action of the $H_1$ on the two Werner-like states, it is observed that for the first input state $\sigma_{AB}^{(1)}$ QFI takes place higher value than second one, that is $F_{\varphi}^{(1)}(\sigma_{AB}^{(1)})>F_{\varphi}^{(1)}(\sigma_{AB}^{(2)})$. Contrary to this situation, for the action of the second Hamiltonian $H_2$, while QFI of the output corresponding to first probe 
state vanishes, $F_{\varphi}^{(2)}(\sigma_{AB}^{(2)})$ attains the maximum value for appropriate choices of the parameters and QCRB is saturated. So, it can be said that QFI strictly depends on the 
choice of the initial probe state and actions of the Hamiltonians. By adjusting the parameters and by adopting the QCRB as a figure of merit, it may also take high values to enhance a better quantum
parameter estimation task.

Secondly, we have investigated the quantum Fisher information dynamics of some different two-qubit input states under the action of the different YBSs. We demonstrated that the QFI about the parameter
$\varphi$ has different behavior for the action of different YBSs or Hamiltonians and introduced the relationship between QFI flow and information to understand the changing trend of QFI. Thus, the QFI
flow of the negative value indicates that the energy and information flow from the system to the environment, corresponding to the QFI of decay whereas the QFI flow of positive value means that
information flows from the environment to the system accounting for the QFI of revival. It is concluded that in this paper, the Hamiltonians constructed with YBMs show non-Markovian behavior in certain
ranges of parameters. Also, it may be worth to study the dynamical evolution of the QFI for multiple parameter estimation in the different states. Our studies on this issue are in progress.

\begin{acknowledgements}
This work was supported in part by the Scientific and Technological Research Council of Turkey (TUBITAK).
\end{acknowledgements}


\begin{thebibliography}{}
\bibitem{Helstrom} Helstrom, C.W.: Quantum Detection and Estimation Theory, Academic Press, New York (1976)
\bibitem{Holevo2001} Holevo, A.S.: Statistical Structure of Quantum Theory. Lect. Not. Phys. 61, Springer, Berlin (2001)
\bibitem{Petz2008} Petz, D.: Quantum information theory and quantum statistics. Springer, Berlin, Heidelberg (2008)
\bibitem{Fisher1925} Fisher, R.A.: Theory of Statistical Estimation. Proc. Camb. Phil. Soc. {\bf 22}, 700-725 (1925)
\bibitem{BN2000} Barndorff-Nielsen O.E., Gill, R.D.: Fisher information in quantum statistics. J. Phys. A {\bf 33}, 4481 (2000)
\bibitem{Paris2009} Paris, M.G.A.: Quantum estimation for quantum technology. Int. J. Quantum Inform. {\bf 07}, 125 (2009)
\bibitem{Toth2014} Tóth, G., Apellaniz, I.: Quantum metrology from a quantum information science perspective. J. Phys. A: Math. Theor. {\bf 47}, 424006 (2014)
\bibitem{Gio1} Giovannetti, V., Lloyd, S., Maccone, L.: Quantum-enhanced measurements: beating the standard quantum limit. Science {\bf 306}, 1330-1336 (2004)
\bibitem{Gio2} Giovannetti, V., Lloyd, S., Maccone, L.: Quantum Metrology. Phys. Rev. Lett. {\bf 96}, 010401 (2006)
\bibitem{Gio3} Giovannetti, V., Lloyd, S., Maccone, L.: Advances in Quantum Metrology. Nature Photon. {\bf 5}, 222-229 (2011)

\bibitem{Kitaev} Kitaev, A.Y.: Fault-tolerant quantum computation by anyons. Ann. Phys. {\bf 303}, 2-30 (2003)
\bibitem{Kauffman} Kauffman, L.H., Lomonaco, S.J. Jr.:  Braiding operators are universal quantum gates. New J. Phys. {\bf 36}, 134 (2004)
\bibitem{Franko} Franko, J.M., Rowell, E.C., Wang, Z.: Extraspecial 2-groups and images of braid group representations. J. Knot Theory Ramif. 15, 413 (2006)
\bibitem{Zhang1} Zhang, Y., Kauffman, L.H., Ge, M.L.: Universal quantum gate, Yang-Baxterization and Hamiltonian. Int. J. Quant. Inf. {\bf 3}, 669 (2005)
\bibitem{Zhang2} Zhang, Y., Ge, M.L.: GHZ states, almost-complex structure and Yang-Baxter equation. Quant. Inf. Proc. {\bf 6}, 363 (2007)
\bibitem{Zhang3} Zhang, Y., Rowell, E.C., Wu, Y.S., Wang, Z.H., Ge, M.L.: From extraspecial twogroups to GHZ states. E-print quant-ph/0706.1761 (2007)
\bibitem{Chen1} Chen, J.L., Xue, K., Ge, M.L.: Braiding transformation, entanglement swapping, and Berry phase in entanglement space. Phys. Rev. A {\bf 76}, 042324 (2007)
\bibitem{Chen2} Chen, J.L., Xue, K., Ge, M.L.: Berry phase and quantum criticality in Yang-Baxter systems. Ann. Phys. {\bf 323}, 2614 (2008)
\bibitem{Chen3} Chen, J.L., Xue, K., Ge, M.L.: All pure two-qudit entangled states generated via a universal Yang-Baxter matrix assisted by local unitary transformations. Chin. Phys. Lett. {\bf 26}, 080306 (2009)
\bibitem{Brylinski} Brylinski, J.L., Brylinski, R.: Universal quantum gates. In: Brylinski, R., Chen, G. (eds.) Mathematics of Quantum Computation, Chapman Hall/CRC Press, Boca Raton (2002)
\bibitem{Wang} Wang, G., Xue, K., Wu, C., Liang, H., Oh, C.H.: Entanglement and Berry phase in a new Yang-Baxter system. J. Phys. A Math. Theor. {\bf 42}, 125207 (2009)
\bibitem{Jones} Jones, V.F.R.: Baxterization. Int. J. Mod. Phys. A {\bf 6}, 2035-2043 (1991)
\bibitem{Ge1991} Ge, M.L., Xue, K., Wu, Y-S.: Explicit trigonometric Yang-Baxterization. Int. J. Mod. Phys. A {\bf 6}, 3735 (1991)
\bibitem{Hu2008} Hu, S. W., Xue, K., Ge, M.-L.: Optical simulation of the Yang-Baxter equation. Phys. Rev. A {\bf 78}, 022319 (2008)
\bibitem{Hu2010} Hu, T., Sun, C., Xue, K.: The sudden death of entanglement in constructed Yang-Baxter systems. Quant. Inf. Proc. {\bf 9}, 27-35 (2010)

\bibitem{Li-Luo} Li, N., Luo, S.: Entanglement detection via quantum Fisher information. Phys. Rev. A {\bf 88}, 014301 (2013)
\bibitem{Song} Song, H., Luo. S., Hong. Y.: Quantum non-Markovianity based on the Fisher-information matrix. Phys. Rev. A {\bf 91} 042110 (2015)
\bibitem{Ban} Ban, M.: Quantum Fisher information of a qubit initially correlated with a non-Markovian environment. Quantum Inf. Process. {\bf 14}, 4163-4177 (2015)
\bibitem{Lu2010} Lu, X.M., Wang, X.G., Sun, C.P.: Quantum Fisher information flow and non-Markovian processes of open systems. Phys. Rev. A {\bf 82}, 042103 (2010)
\bibitem{Zhong2013} Zhong, W., Sun, Z., Ma, J., Wang, X-G., Nori, F.: Fisher information under decoherence in Bloch representation. Phys. Rev. A {\bf 87}, 022337 (2013)
\bibitem{Breuer} Breuer, H.-P., Petruccione, F.: The Theory of Open Quantum Systems, Oxford University Press, Oxford (2007)
\bibitem{Alicki} Alicki, R., Lendi, K.: Quantum Dynamical Semigroups and Applications, Springer, Berlin (2007)
\bibitem{Rivas} Rivas, \'{A}., Huelga, S.F., Plenio, M.B.: Quantum non-Markovianity: characterization, quantification and detection, Rep. Prog. Phys. {\bf 77}, 094001 (2014)

\bibitem{Holevo1982} Holevo, A.S., Ballentine, L.E.: \textit{Probabilistic and Statistical Aspects of Quantum Theory}. NORTH HOLLAND, (1982)
\bibitem{Braunstein} Braunstein S.L., Caves, C.M.: Statistical distance and the geometry of quantum states. Phys. Rev. Lett. {\bf 72}, 3439 (1994)
\bibitem{Cramer1946} Cram\'{e}r, H.: \textit{Mathematical Methods of Statistics}. Princeton University Press, Princeton, (1946)
\bibitem{Rao} Rao, C.R.: Information and the accuracy attainable in the estimation of statistical parameters. Bull. Calcutta Math. Soc. {\bf 37}, 81-89 (1945)
\bibitem{Liu1} Liu, J., Jing, X. X, Wang, X. (2013). Phase-matching condition for enhancement of phase sensitivity in quantum metrology. Phys. Rev. A {\bf 88}, 042316.
\bibitem{Zhang2014} Zhang, Y.M., Li, X.W., Yang, W., Jin. G.R.: Quantum Fisher information of entangled coherent states in the presence of photon loss. Phys. Rev. A {\bf 88}, 043832 (2013)
\bibitem{Liu2} Liu, J., Jing, X.X., Zhong, W., Wang, X.G.: Quantum Fisher Information for Density Matrices with Arbitrary Ranks. Commun. Theor. Phys. {\bf 61}, 45-50 (2014)
\bibitem{Jing} Jing, X. X., Liu, J., Zhong, W., Wang, X.G.: Quantum Fisher Information of Entangled Coherent States in a Lossy Mach-Zehnder Interferometer. Commun. Theor. Phys. {\bf 61}, 115-120 (2014)
\bibitem{Liu3} Liu, J., Yuan, H., Lu, X.M., Wang, X.G.: Quantum Fisher information matrix and multiparameter estimation. arXiv:1907.08037 (2019)
\bibitem{Boixo} Boixo, S., Flammia, S.T., Caves, C.M., Geremia, J.M.: Generalized Limits for Single-Parameter Quantum Estimation. Phys. Rev. Lett. {\bf 98}, 090401 (2007)
\bibitem{Liu4} Liu, J., Jing, X.X., Wang, X.G.: Quantum metrology with unitary parametrization processes. Sci. Rep. {\bf 5}, 8565 (2015)
\bibitem{Taddei} Taddei, M.M., Escher, B.M., Davidovich, L., de Matos Filho, R.L.: Quantum Speed Limit for Physical Processes. Phys. Rev. Lett. {\bf 110}, 050402 (2013)
\bibitem{Lin2008} Lin, D., Liu, Y., Zou, H.-M.: Modulating quantum Fisher information of qubit in dissipative cavity by coupling strength. Chin. Phys. B {\bf 27(11)}, 110303 (2018)
\bibitem{Abdel-Khalek} Abu-Zinadah, H.H., Abdel-Khalek, S.: Fisher information and quantum state estimation of two-coupled atoms in presence of two external magnetic fields. Results in Physics {\bf 7}, 4318-4323 (2017)
\bibitem{Haikka} Haikka, P., Goold, J., McEndoo, S., Plastina, F., Maniscalco, S.: Non-Markovianity, Loschmidt echo, and criticality: A unified picture, Phys. Rev. A {\bf 85}, 060101(R) (2012)
\bibitem{Hao1} Hao, X., Wu, W., Zhu, S.: Nonunital non-Markovian dynamics induced by a spin bath, interplay of quantum Fisher information, arXiv:1311.5952 (2013)
\bibitem{Hao2} Hao, X., Tong, N-H., Zhu, S.: Dynamics of the quantum Fisher information in a spin-boson model, J. Phys. A: Math. Theor. {\bf 46}, 355302 (2013)

\bibitem{Yang} Yang, C.N.: Some exact results for the many-body problem in one dimension with repulsive delta-function interaction. Phys. Rev. Lett. {\bf 19}, 1312-1315 (1967);
Yang, C.N.: $S$ matrix for the one-dimensional $N$-body problem with repulsive or attractive $\delta$-function interaction. Phys. Rev. {\bf 168}, 1920 (1968)
\bibitem{Baxter} Baxter, R.J.: Partition function of the Eight-Vertex lattice model. Ann. Phys. {\bf 70}, 193-228 (1972); Baxter, R.J.:
Exactly Solved Models in Statistical Mechanics. Academic Press, London (1982)
\bibitem{Jimbo1989} Jimbo, M.: Introduction to the Yang-Baxter equation. International Journal of Modern Physics A {\bf 4 (15)}, 3759-3777 (1989)
\bibitem{Temperley} Temperley, H.N.V., Lieb, E.H.: Relations between the `percolation' and `colouring' problem and other graph-theoretical problems associated with regular planar lattices:
some exact results for the `percolation' problem. Proc. Roy. Soc. London A {\bf 322}, 251-280 (1971)
\bibitem{Hu2007} Hu, S.W., Hu, M.G., Xue, K., Ge, M.L.: Linear optics implementation for Yang-Baxter equation. arXiv: 0711.4703v2 (2007)
\bibitem{Sun2009} Sun, C., Hu, T., Wu, C., Xue, K.: Thermal entanglement in the systems constructed from the Yang-Baxter R-matrix. Int. J. Quant. Inf. {\bf 7 (5)}, 879-889 (2009)
\bibitem{NC} Nielsen, M., Chuang, I.L.: Quantum Computation and Quantum Information, $10th$ Anniversary Edition. Cambridge University Press, Cambridge (2010)
\bibitem{Luo} Luo, S.: Quantum discord for two-qubit systems. Phys. Rev. A {\bf 77}, 042303 (2008)

\end{thebibliography}


\end{document}